# A Way to Design Diaelectric Materials

Yi-Neng Huang[1) 2)†] Xing-Yu Zhao[1)2)] Li-Li Zhang[1)2)] Xin-Ru Huang[3)]

1) National Laboratory of Solid State Microstructures, School of Physics, Nanjing University, Nanjing, 210093, China
2) Xinjiang Laboratory of Phase Transitions and Microstructures in Condensed Matters, College of Physical Science and Technology, Yili Normal University, Yining, 835000, China
3) Department of Physics, Emory University, Atlanta, 30322, USA

**ABSTRACT**

Considering the coupling between electric polarization and crystal lattice in ferroelectrics, the authors propose a new molecular field theory. It not only includes the classical Weiss Molecular field (WMF), but also the spontaneous distortion inducted molecular field (SDIMF). This theory tells us that diaelectric state will appear in ferroelectrics when the direction of SDIMF is opposite to WMF and its strength is strong enough. In other words, we found a way to design diaelectric materials (diaelectrics). Besides, the theory also predicts the appearance of large static dielectric constant state in ferroelectric phase, when the SDIMF and WMF are in opposite direction but the SDIMF strength is relative small. This is definitely valuable in mentoring the design of new relevant materials.

## 1. Introduction

Diamagnetism means that the static susceptibility is smaller than zero, and it is a general property of materials. However, according to the knowledge of the authors, diaelectric materials, whose static polarizability $\chi_s < 0$, has never been discovered or predicted in theory. In this paper, considering the coupling between electric polarization and crystal lattice, we provide a new way to design diaelectric materials (diaelectrics).

## 2. Designing theory of diaelectrics

The coupling between electric polarization and crystal lattice is relatively strong in ferroelectrics, and the spontaneous polarization $P_s$ is always accompanied with spontaneous distortion (changes of lattice parameters and/or symmetry) during the ferroelectric phase transition[1,2], so the distortion along with $P_s$ inevitably leads to the variations of interaction between dipoles[3] or molecular field $E_m$ acting on dipoles[4]. Based on inversion symmetry of $P_s$, here we think about the following $E_m$ related to $P_s$,

$$E_m = \frac{P_s}{N\mu^2}\left[\beta_0 + \beta_2\left(\frac{P_s}{N\mu}\right)^2\right] \quad (1)$$

where $\mu$ and $N$ are the permanent dipole moment and the density per unit volume of dipoles, respectively. The term of $\beta_0$ is the Weiss molecular field (WMF)[4], and that of $\beta_2$ is the spontaneous distortion induced molecular field (SDIMF). Here we define $\gamma \equiv \beta_2/\beta_0$, and it represents the relative strength and direction of SDIMF to WMF.

Supposing there are only two orientation states of dipoles in ferroelectrics, and according to the molecular field of Eq.(1) as well as the Boltzmann principal [1-4], we can get $P_s$ and $\chi_s$ of the system separately,

$$P_s = N\mu \tanh\left[-\frac{\mu E_m}{k_B T}\right] \quad (2)$$

$$\chi_s = \frac{C_w\left[1 - \tanh^2\left(\frac{\mu E_m}{k_B T}\right)\right]}{T - T_\beta\left[1 - \tanh^2\left(\frac{\mu E_m}{k_B T}\right)\right]} \quad (3)$$

† E-mail: ynhuang@nju.edu.cn

where $k_B$ is the Boltzmann constant, $T_\beta \equiv \beta_0/k_B$, $C_w \equiv N\mu^2/(k_B\varepsilon_0)$, and $\varepsilon_0$ is dielectric constant of vacuum.

paraelectric state in ferroelectric phase is not a phase transition [1-3], but a kind of crossover (Fig.1a and the inset of Fig.2).

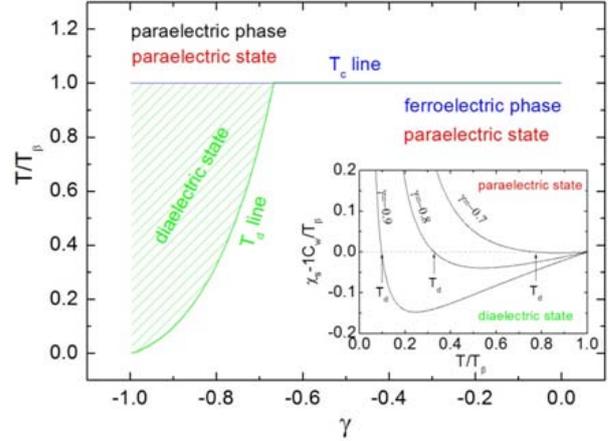

Fig.2 Calculated $T_d$ and $T_c$ vs $\gamma$ by the theory in this paper. The inset shows $\chi_s^{-1}$ of ferroelectric phase vs T when $\gamma$=-0.7、-0.8、-0.9.

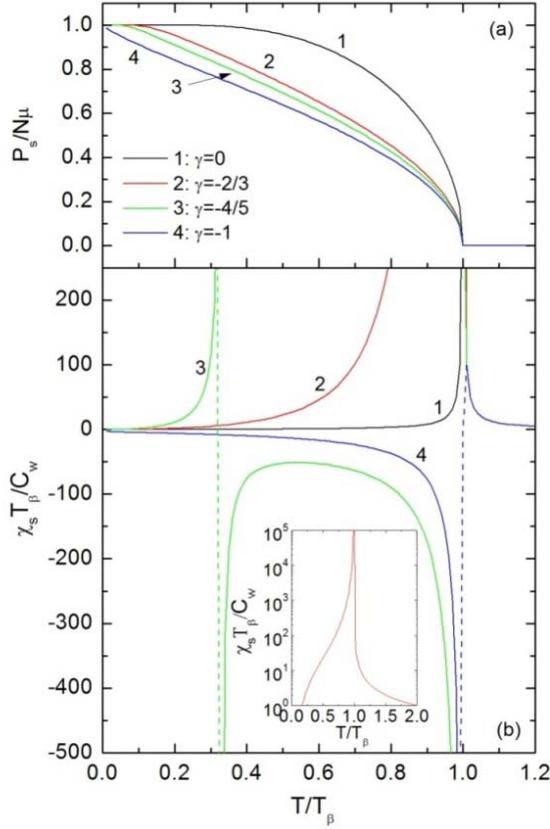

Fig.1 Calculated $P_s$ (a) and $\chi_s$ (b) of the system vs T for a set of $\gamma$ values by the theory in this paper. The inset shows $\chi_s$ vs T when $\gamma$=-2/3.

## 3. Theoretical results and discussions

Fig.1 shows the calculated results of $P_s$ and $\chi_s$ changes with T when $\gamma$=0, -2/3, -4/5, and -1 by the present theory. From Fig.1a, it can be seen that $P_s$ begins to increase continuously from $T_\beta$ during the cooling process. The continuous spontaneous polarization means that continuous or 2$^{nd}$ order ferroelectric phase transition happens in the system [1-3], and the transition temperature $T_c = T_\beta$.

It is particularly deserving to point out that, when $-1 < \gamma < -2/3$, the system exhibits $\chi_s < 0$ between $T_c$ and a specific temperature $T_d$ (defined as diaelectric temperature here) as shown in Fig.1b and the inset of Fig.2, and we call it diaelectric state. It is noticed that the transition between diaelectric state and

Fig.2 shows the calculated $T_d$ vs $\gamma$ and the existing zone of diaelectric state by the present theory. Evidently, it must be ensured that the direction of SDIMF to WMF is opposite ($\gamma < 0$) and its strength is large enough when we design ferroelectrics.

It is worth noting that when $-2/3 \le \gamma < 0$, $\chi_s$ has a relative large value in wide temperature zone in ferroelectric phase, and this is definitely guiding significant in designing relevant materials.

**Acknowledgement**: This work is subsidized by national 973 program of China (Grant No. 2012CB821503); Xinjiang high-tech development project (Grant No. 200916126), Xinjiang, China; and Xinjiang natural fund (Grant Nos. 200821104, 200821184), Xinjiang, China.